\documentclass[a4paper]{jpconf}

\usepackage{graphicx}
\usepackage[T1]{fontenc}
\usepackage[english]{babel}
\usepackage{color}
\usepackage{amsmath}
\usepackage[numbers]{natbib}
\usepackage[varg]{txfonts}
\usepackage{hyperref}

\newcommand{\code}[1]{\textsc{#1}}
\newcommand{\incl}{\code{INCL}}
\newcommand{\geminipp}{\code{GEMINI++}}

\begin{document}

\title{Simultaneous fitting of statistical-model parameters to symmetric and
  asymmetric fission cross sections}

\author{D~Mancusi$^1$, R~J~Charity$^2$ and J~Cugnon$^3$}

\address{$^1$ CEA, Centre de Saclay, Irfu/Service de Physique Nucléaire, F-91191
  Gif-sur-Yvette, France}

\address{$^2$ Department of Chemistry, Washington University, St.\ Louis,
  Missouri 63130, USA}

\address{$^3$ University of Li\`ege, AGO Department, all\'ee du 6 Ao\^ut 17,
  b\^at. B5, B-4000 Li\`ege 1, Belgium}

\ead{davide.mancusi@cea.fr}

\begin{abstract}
  The de-excitation of compound nuclei has been successfully described for
  several decades by means of statistical models. However, accurate predictions
  require some fine-tuning of the model parameters. This task can be simplified
  by studying several entrance channels, which populate different regions of the
  parameter space of the compound nucleus.

  Fusion reactions play an important role in this strategy because they minimise
  the uncertainty on the entrance channel by fixing mass, charge and excitation
  energy of the compound nucleus. If incomplete fusion is negligible, the only
  uncertainty on the compound nucleus comes from the spin distribution. However,
  some de-excitation channels, such as fission, are quite sensitive to
  spin. Other entrance channels can then be used to discriminate between
  equivalent parameter sets.

  The focus of this work is on fission and intermediate-mass-fragment emission
  cross sections of compound nuclei with $70\lesssim A\lesssim240$. The
  statistical de-excitation model is \geminipp. The choice of the observables is
  natural in the framework of \geminipp, which describes fragment emission using
  a fission-like formalism. Equivalent parameter sets for fusion reactions can
  be resolved using the spallation entrance channel. This promising strategy can
  lead to the identification of a minimal set of physical ingredients necessary
  for a unified quantitative description of nuclear de-excitation.
\end{abstract}

\section{Introduction}

The de-excitation of an excited nucleus is a qualitatively well-understood
phenomenon which is often described by means of statistical models. However,
such models contain a great deal of free parameters and ingredients that are
often underconstrained by the available experimental data. Quantitatively
accurate predictions usually require some tuning of the model parameters.

The fusion entrance channel is a particularly powerful tool to explore the
sensitivity of the de-excitation model to the compound-nucleus parameters (mass,
charge, excitation energy and spin); if the cross sections for incomplete fusion
and pre-equilibrium emission are negligible with respect to the fusion cross section
for a given projectile-target combination, the compound nucleus can essentially
be regarded as having a fixed mass, charge and total excitation energy
(intrinsic plus collective), thereby fixing three of the four parameters that
describe it. The requirement of complete fusion, however, puts an upper limit on
the energy of the projectile and, thus, on the excitation energies that can be
studied with this method. Because of this and other similar limitations on the
entrance channel, one is actually able to construct different parameter sets
that can describe the same experimental data to a similar degree of accuracy; in
this sense, statistical de-excitation models contain partly degenerate
ingredients, and that limits their predictive power.

Part of the degeneracy can be removed by performing simultaneous fits to
heterogeneous data sets. For example, one can try to explore diverse regions of
the compound-nucleus parameter space by studying different reaction entrance
channels. The present work combines the fusion and the spallation entrance
channels for the study of fission and emission of intermediate-mass fragments
(IMFs). The goal is to put more stringent constraints on the de-excitation-model
parameters than those that would be obtained from the separate study of fusion-
and spallation-induced de-excitation chains.

\section{Tools}

This work focuses on the \geminipp\ nuclear de-excitation model
\cite{charity-gemini++}. One of the most prominent features of \geminipp\ is
that it accurately models changes in orbital and intrinsic angular momentum of
the de-excitation products along the de-excitation chain. This is particularly
important for the study of fission and IMF emission, which are quite sensitive
to the spin of the mother nucleus.

For nuclides above the Businaro-Gallone point, the ridge of conditional saddle
points as a function of the asymmetry of the split exhibits a minimum around
symmetric splitting and two local maxima on either side (apart from local
variations due to structure effects). For such systems, \geminipp\ adopts a
global description of fission. The statistical width of the process is computed
using a Bohr-Wheeler-type formalism, with barriers taken from Sierk's
finite-range calculations \cite{sierk-asyBar,carjan-asyBar}. In addition,
several corrections are possible within the framework of the model: (a)
different level-density parameters at the saddle point and in the ground state,
(b) a constant shift of the Sierk barrier heights, (c) overall scaling of the
fission width, and (d) explicit treatment of the tilting degree of freedom at
saddle \cite{lestone-tilting}. This establishes the free ingredients of our
fission model. The scission mass and charge distributions are taken from Rusanov
\etal's systematics \cite{rusanov-mass}.

\geminipp\ also considers the emission of fragments with $3<A<A_\text{IMF}$,
where $A_\text{IMF}$ is the fragment mass corresponding to the first maximum in
the ridge of conditional saddle points. This process is also described by a
transition-state formalism \cite{moretto-binarydecay}, with explicitly
singled-out mass- and charge-asymmetry degrees of freedom at saddle. Given the
formal similarity, the IMF-emission model includes the same free ingredients as
the fission model. Finally, the emission of nucleons and light clusters
($A\leq3$) is described by the Hauser-Feshbach evaporation formalism
\cite{hauser-evaporation}.

\subsection{Models for the entrance channel}

Besides the de-excitation model, the proposed task requires models for the
reaction entrance channels. For fusion, we limit our study to incident energies
lower than about 10~$A$MeV, where incomplete fusion and pre-equilibrium should
be negligible; thus, we only need to specify the spin distribution of the
compound nucleus. We assume the following roughly triangular shape:
\[
\sigma_\text{fus}(J) = \pi \lambdabar^2 (2J+1) {\left[1+\exp \left(
      \frac{J-J_0}{\Delta J}\right)\right]}^{-1}\text,
\]
where $J_0$ determines the maximum spin value and $\Delta J$ plays the role
of a smooth cutoff. The $J_0$ parameter is fixed from the total fusion cross
section
\[
\sigma_\text{fus} = \sum_{J=0}^{\infty} \sigma_\text{fus}(J)\text,
\]
while $\Delta J$ is set to values from 3 to 10~$\hbar$, with the larger values
associated with the heavier projectiles. For the reactions for which we present
IMF data, the fusion cross sections have not been measured and the Friction model
\cite{gross-friction} or the Extra-Push model \cite{swiatecki-extra_push} were used
to calculate both the cross sections and maximum spin values.

The entrance channel for spallation reactions is described by the Li\`ege
Intranuclear-Cascade model (\incl) \cite{boudard-incl}. In this framework, the
high-energy incident nucleon initiates an avalanche of binary nucleon-nucleon
collisions within the target nucleus, which can lead to the emission of a few
nucleons and possibly pions. The cascade is stopped when the cascade remnant
shows signs of thermalisation. This provides the entry point for the \geminipp\
de-excitation chain. A more comprehensive description of the latest \incl\
developments has been recently published \cite{cugnon-incl45_nd2010}. One should
stress here that the \incl\ model does have internal parameters, but they have
been either taken from known phenomenology (e.g.\ the parameters describing
nuclear density distributions) or fixed once and for all (e.g.\ the parameters
connected with the description of Pauli collision blocking). Thus, the present
work only focuses on the adjustment of the \geminipp\ side of the reaction
model.

The validity of the \incl\ model in the 50-MeV to 3-GeV incident-energy range
has been extensively demonstrated by the recent ``Benchmark of Spallation
Models'' \cite{leray-intercomparison,intercomparison-website}, sponsored by
IAEA. We assume that \incl\ provides an accurate description of the initial
stage of spallation reactions within the energy range above. Above 3~GeV, the
model is known to be less reliable \cite{pedoux-pions}. This limits the pool of
experimental data that can be considered if we require that the entrance-channel
model should not introduce considerable uncertainty on the model predictions.

\subsection{Complementarity of fusion and spallation}

\begin{figure}
  \centering
  \includegraphics*[width=0.6\linewidth]{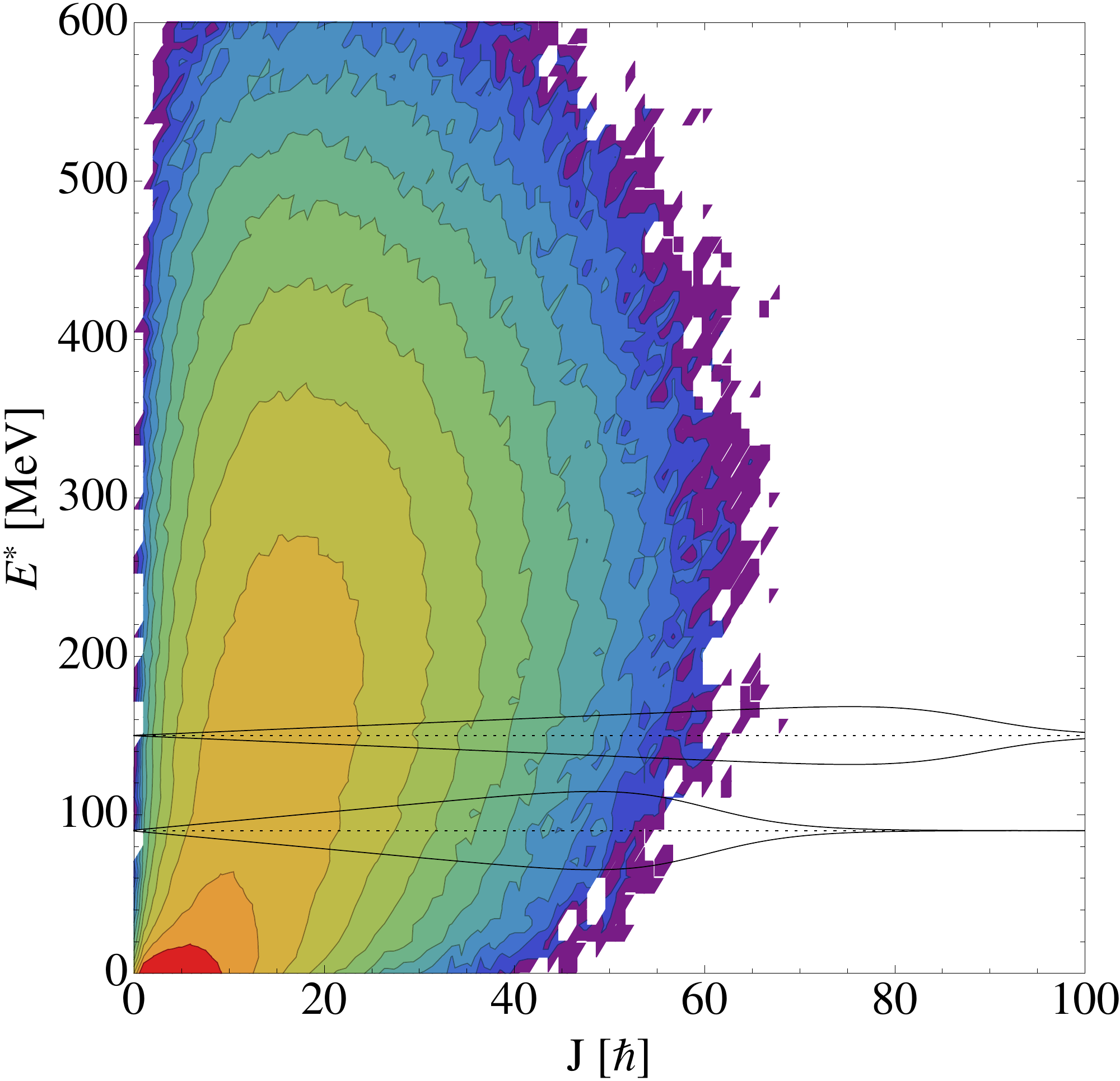}
  \caption{Comparison of the distributions of excitation energy and spin
    populated in the $^{19}$F+$^{181}$Ta$\rightarrow^{200}$Pb fusion reaction
    for $E^*={}$90,
    150~MeV (horizontal lines) with the \incl\ prediction for the 1-GeV
    \textit{p}+$^{208}$Pb spallation reaction (contours, logarithmically
    spaced).}
  \label{fig:ejmap}
\end{figure}

Figure~\ref{fig:ejmap} illustrates how spallation and fusion reactions
efficiently complement each other in probing the parameter space of thermalised
nuclei. Spallation reactions produce broad distributions of excited nuclei,
whose projection on the spin/excitation-energy plane is represented by the
coloured contours. Rather high excitation energies can be realised, but spin is
limited to a few tens of $\hbar$. This complements well the limitations of
fusion reactions, which are represented by the horizontal shapes. The width of
the shapes is proportional to the spin distributions of the
$^{19}$F+$^{181}$Ta$\rightarrow^{200}$Pb fusion reaction, for two different
excitation energies.

\section{Results for fission}
\label{sec:fission}

\begin{figure}[t]
  \begin{minipage}[t]{0.47\linewidth}
    \includegraphics[width=\linewidth]{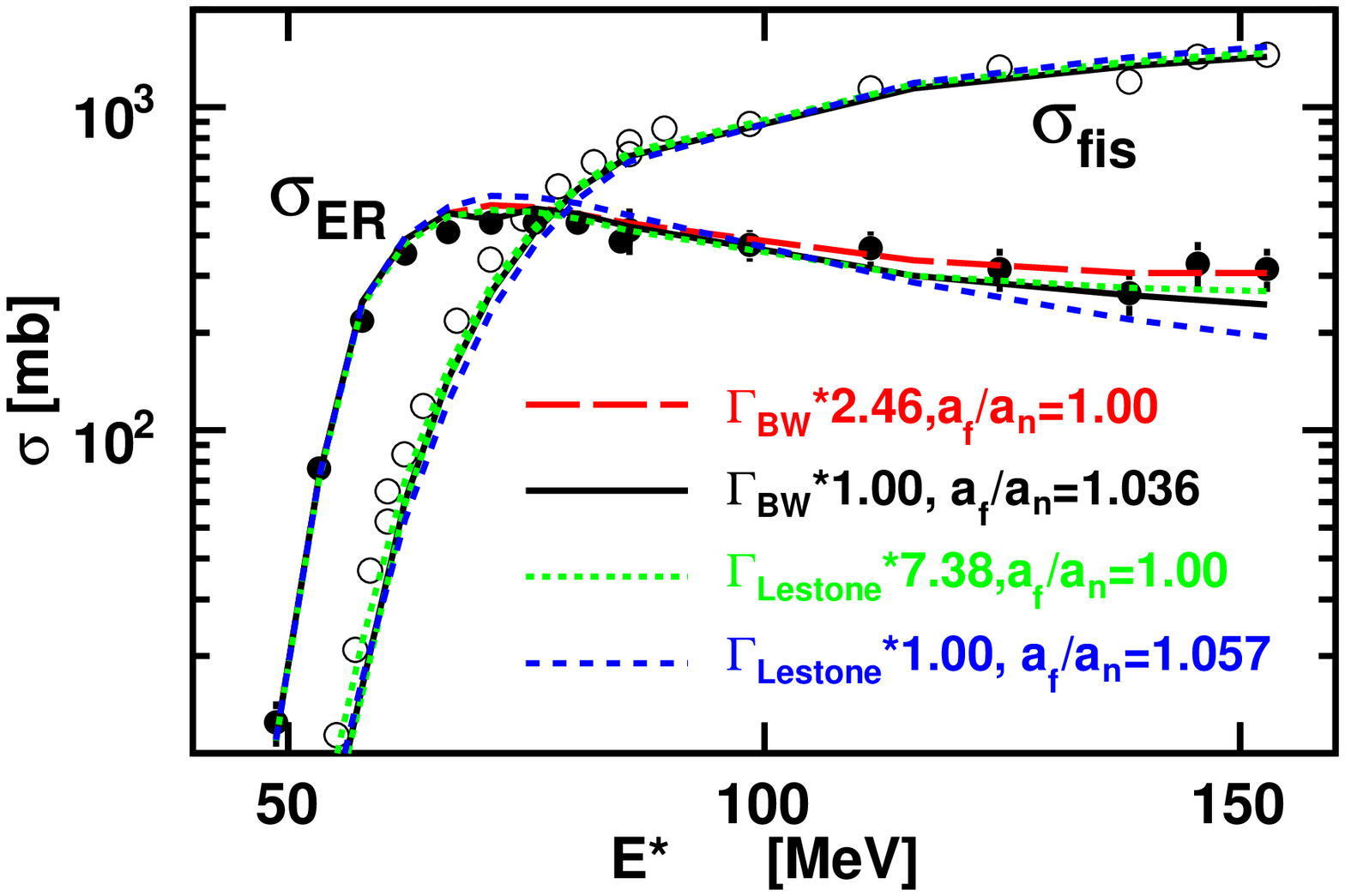}
    \caption{Experimental \protect\cite{Hinde82,Caraley00} and calculated
      \geminipp\ predictions for evaporation-residue and fission excitation
      functions for the $^{19}$F+$^{181}$Ta reaction.}
    \label{fig:fitPb200}
  \end{minipage}\hspace{0.06\linewidth}%
  \begin{minipage}[t]{0.47\linewidth}
    \includegraphics*[width=\linewidth]{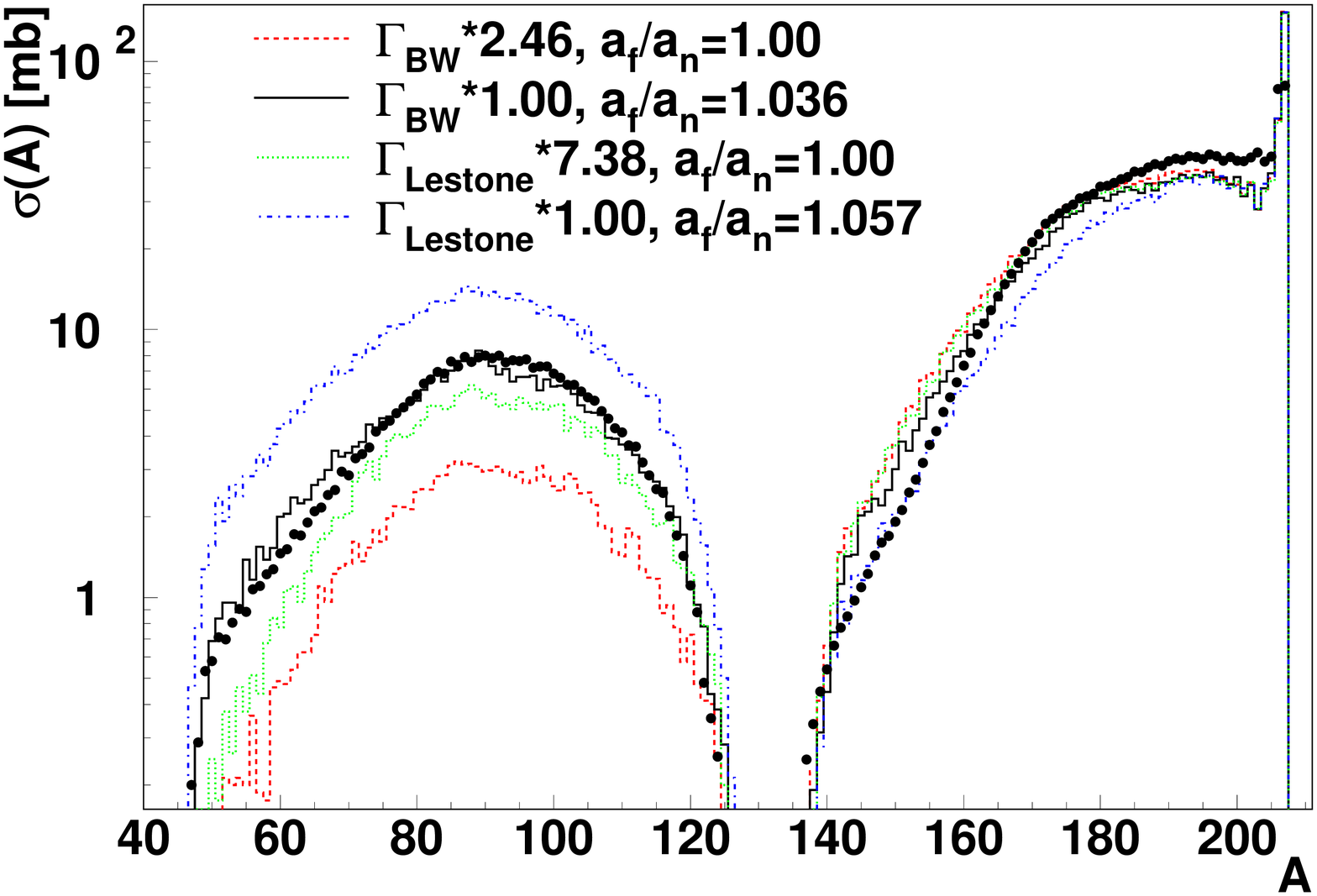}
    \caption{Experimental \protect\cite{enqvist-lead} and calculated residual mass
      distributions for the 1-GeV \textit{p}+$^{208}$Pb reaction.}
    \label{fig:fitpPb}
  \end{minipage}
\end{figure}

We first discuss fusion-fission and spallation-fission calculations for compound
nuclei of similar mass and charge. Figure~\ref{fig:fitPb200} shows the result of
four fits to fusion-fission data: here $\Gamma_\text{BW}$ and
$\Gamma_\text{Lestone}$ indicate calculations performed without or with
Lestone's tilting correction, respectively; a global scaling factor is applied
in some parameter sets; and $a_\text{f}/a_\text{n}$ represents the ratio of the
level-density parameters at saddle and in the ground-state (assumed to be a
constant). The degeneracy of the four parameter sets is clearly
illustrated. However, the application of the same parameter sets to
spallation-fission reactions largely lifts the degeneracy for this observable,
as shown in Figure~\ref{fig:fitpPb}.

The combined fusion/spallation approach allowed us to construct predictive
parameter sets for fission and evaporation-residue excitation curves in the
compound-nucleus mass
range $155\lesssim A\lesssim225$ \cite{mancusi-gemini++_fission}. The agreement
with experimental data for both types of reactions was in general very good,
apart from some overestimation of fission cross sections for the lightest
compound nuclei ($A<170$).

\section{Results for IMF emission}

We now proceed to illustrate the application of the same strategy to fragment
charge distributions from non-fissile compound nuclei, which cover
IMF-production cross sections. Given the paucity of available data, we also
considered experimental data for reactions above 10 $A$MeV incident energy, for
which incomplete fusion might not be negligible. In all cases, however, the
authors of the experimental papers state that the incomplete-fusion component
has been properly subtracted.

\subsection{Fusion reactions}

\begin{figure}[t]
  \centering
  \includegraphics[width=0.7\linewidth]{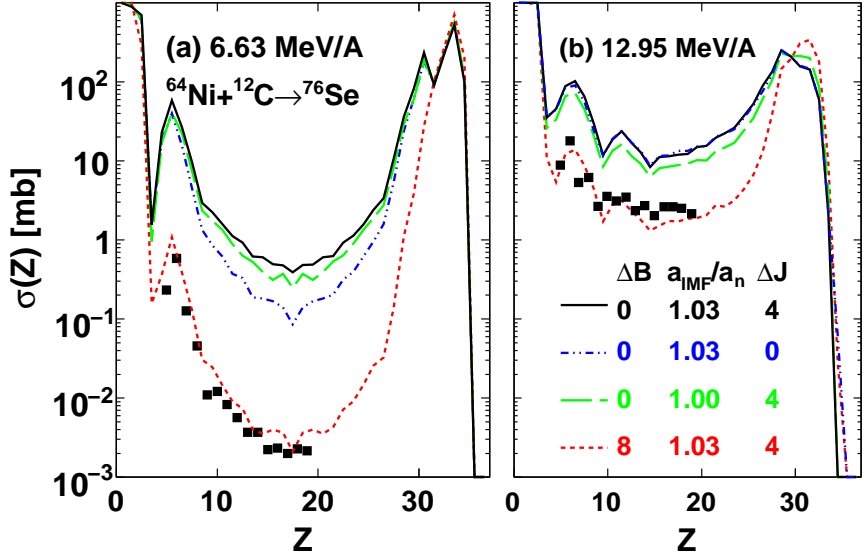}
  \caption{Comparison of the experimental charge distribution for the (a)
    $E/A=6.63$~MeV and (b) $E/A=12.95$~MeV $^{64}$Ni+$^{12}$C fusion reaction
    with calculation with the indicated input parameters. The fusion cross
    section is given by the Friction model.}
  \label{fig:fusion_imf}
\end{figure}

Figure~\ref{fig:fusion_imf} shows the charge distribution of fragments obtained
from $^{58}$Ni+$^{12}$C$\to{}^{76}$Se
fusion at 6.63 and 12.95~$A$MeV incident energy. The figure illustrates the
sensitivity of the calculation results to three parameters: a constant shift of
the Sierk IMF barriers ($\Delta B$), the saddle-to-ground-state ratio of
level-density parameters ($a_\text{IMF}/a_\text{n}$, analogous to the
$a_\text{f}/a_\text{n}$ parameter for fission) and the diffuseness parameter of
the spin distribution ($\Delta J$). IMF yields from fusion show great
sensitivity to the barrier height, which is expected because the
compound-nucleus nuclear temperature ($T\sim{}$a
few MeV) is much smaller than the typical IMF barrier height ($B\sim{}$a
few tens of MeV) and the transition rate scales approximately as
$\exp(-B/T)$. For the same reason, IMF yields from fusion are relatively
insensitive to the small variation of the $a_\text{IMF}/a_\text{n}$ ratio, which
determines the temperature $T$. The diffuseness of the spin distribution also
has some effect on the IMF yields, especially at low excitation energy.  We
observe that the experimental data for this system can be satisfactorily
reproduced only by adding a shift of 8.5~MeV to Sierk's barriers. Moreover, the
data do not show any clear need for an asymmetry dependence of the barrier
shift.

\begin{figure}[t]
  \includegraphics[width=\linewidth]{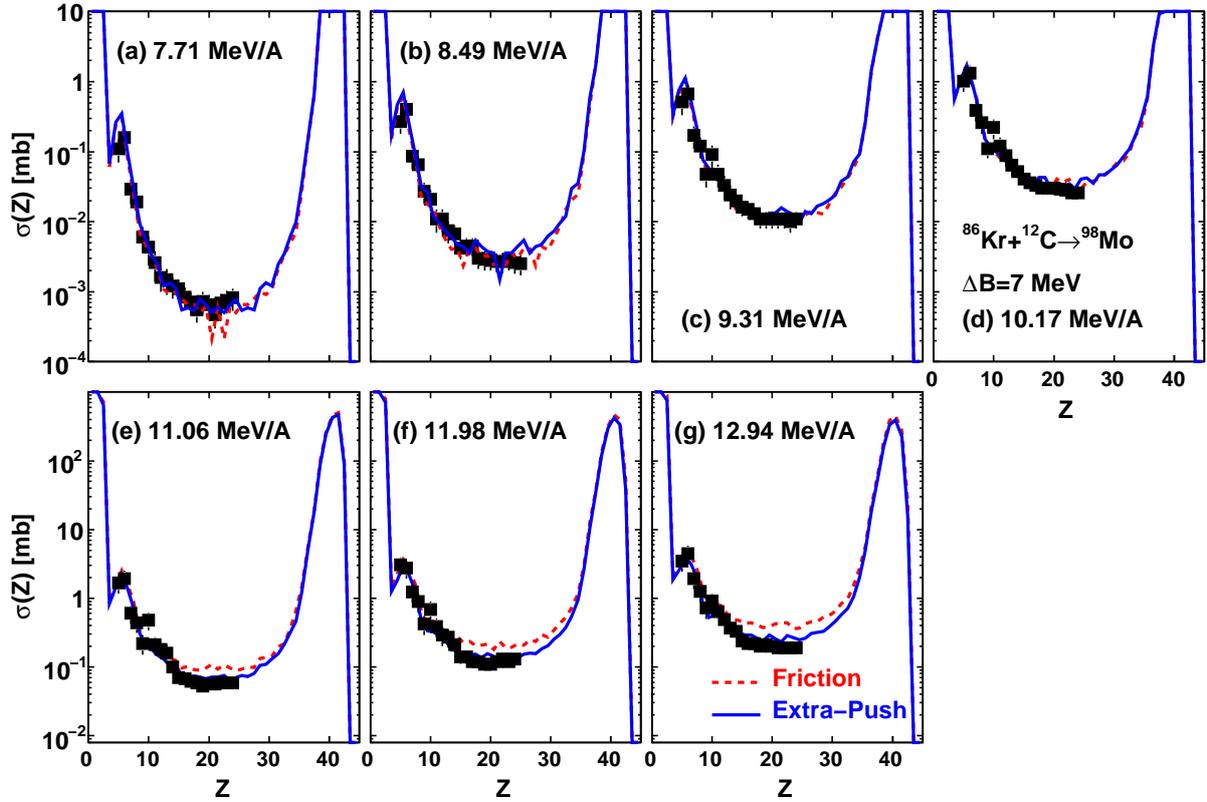}
  \caption{Comparison of experimental and fitted charge distributions
    for the $^{86}$Kr+$^{12}$C reaction.}
  \label{fig:mo98fit}
\end{figure}

Figure~\ref{fig:mo98fit} shows how IMF-production cross sections from another
fusion reaction ($^{86}$Kr+$^{12}$C$\to{}^{98}$Mo)
can be accurately described at several excitation energies by applying an
asymmetry-independent shift of 7~MeV to the Sierk barriers. The insensitivity of
the model predictions to the fusion-cross-section model is also illustrated by
the comparison of the Friction and Extra-Push models.

Following this approach, we have fitted barrier shifts to all the available data
sets \cite{fan-se,delis-br75,jing-mo,sobotka-decay,charity-niobium}.  All charge
distributions were fitted with $a_\text{IMF}/a_\text{n}$=1.03 and by varying the quantity
$\Delta B$. The choice of $a_\text{IMF}/a_\text{n}$ will be discussed below
(Sec.~\ref{sec:spal}), but it is generally not very important for the considered
fusion reactions.

\begin{figure}[t]
  \begin{minipage}[t]{0.47\linewidth}
    \centering
    \includegraphics[width=\linewidth]{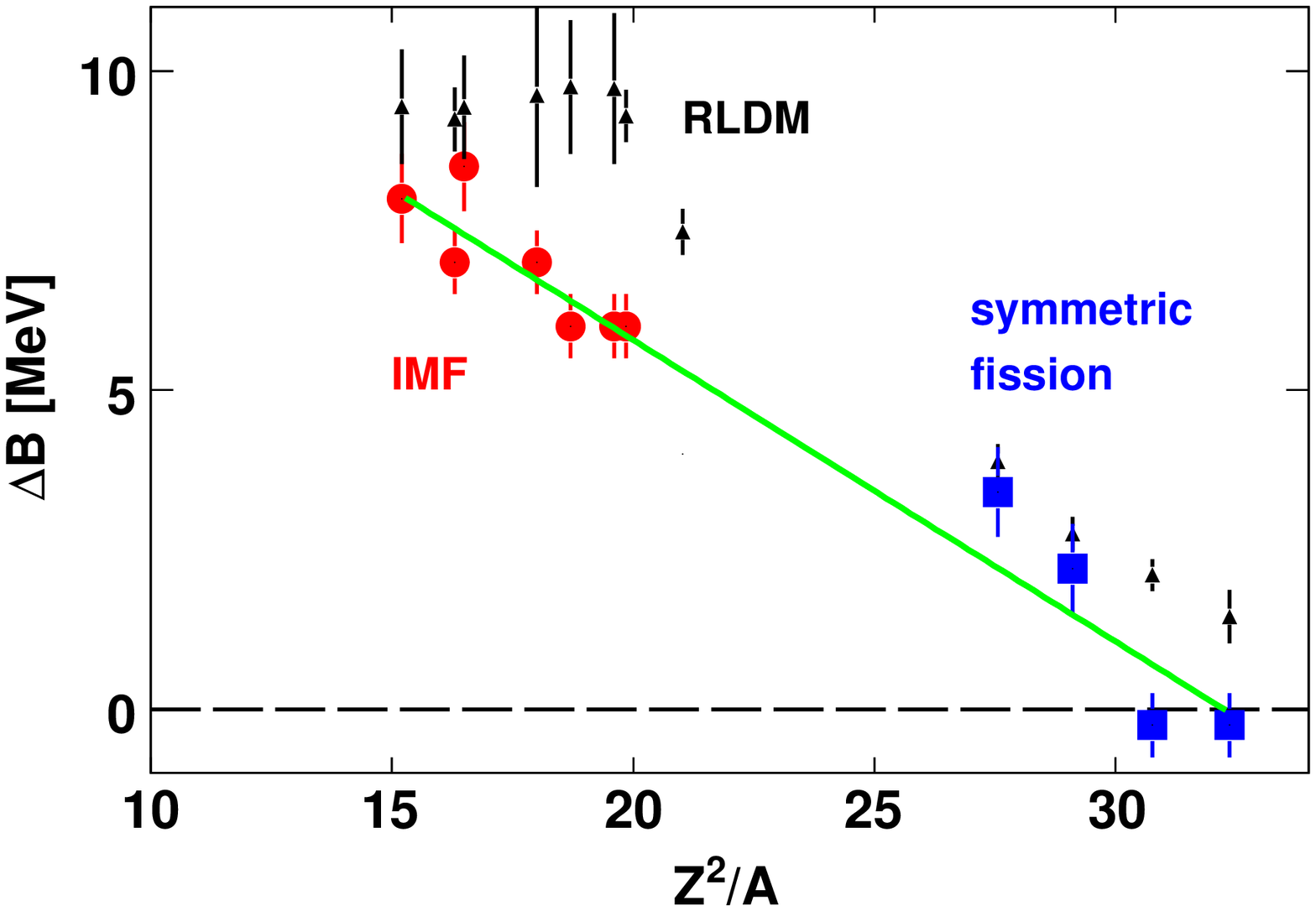}
    \caption{Shifts to the finite-range barriers $\Delta B$ needed to fit data,
      plotted as a function of the $Z^2/A$ of the compound nucleus.}
    \label{fig:shift4}
  \end{minipage}\hspace{0.06\linewidth}%
  \begin{minipage}[t]{0.47\linewidth}
    \includegraphics[width=\linewidth]{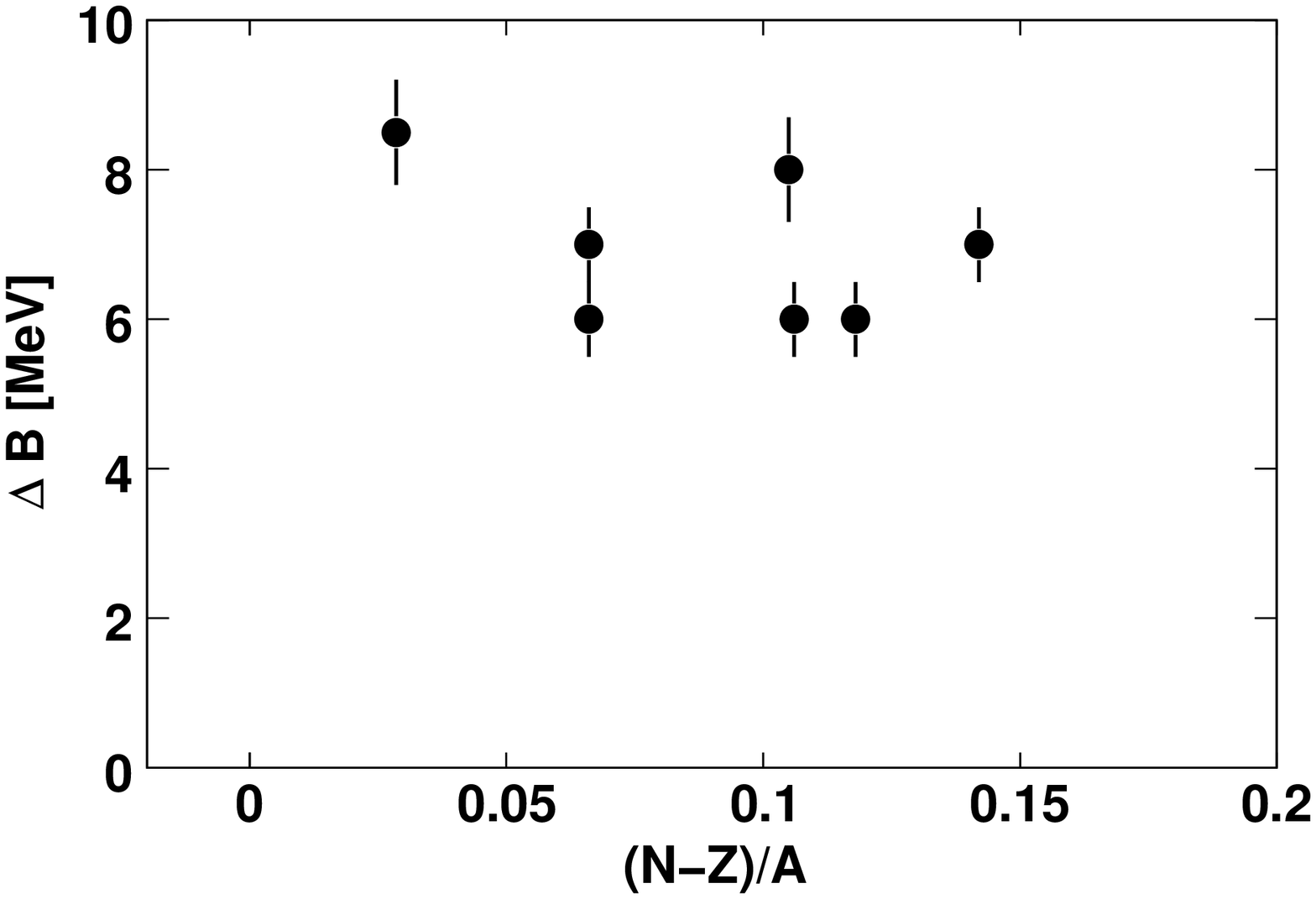}
    \caption{Shifts to the finite-range barriers $\Delta B$ needed to fit the
      experimental IMF charge distributions, plotted as a function of the
      asymmetry parameter.}
    \label{fig:shiftI}
  \end{minipage}
\end{figure}

The fitted barrier shifts correlate well with the fissility parameter $Z^2/A$ of
the compound nucleus, as shown in Figure~\ref{fig:shift4}. The circular data
points are from fits to IMF charge distributions.

As mentioned above (Sec.~\ref{sec:fission}), fission cross sections for the
lightest compound nuclei were slightly overestimated by our global fission fits
\cite{mancusi-gemini++_fission}. The quality of the fit to fission excitation
curves can be improved by increasing the fission barriers by a few MeV. The
best-fit shifts are represented by the square data points on
Figure~\ref{fig:shift4} and seem to align with the trend shown by the IMF charge
distributions. The line shows a linear fit to the red and blue data points.

The physical interpretation of the barrier shift is unclear. The triangular data
points on Figure~\ref{fig:shift4} represent differences between Sierk's
finite-range barriers and the Rotating Liquid-Drop barriers \cite{cohen-rldm},
which are mainly due to the finite-range and surface-diffuseness corrections
included in Sierk's model. The dependence on $Z^2/A$ is similar to our barrier
shifts. This might indicate that Sierk's model overestimates the extent of the
correction.

One can also try to explain the barrier shift as an effect of the deformation
dependence of the Wigner energy of the mother nucleus
\cite{myers-wigner_energy}. However, if this interpretation were correct, the
barrier shifts should show some dependence on $|N-Z|/A$, the asymmetry parameter
of the compound nucleus, with a minimum close to symmetry. No clear trend
appears on Figure~\ref{fig:shiftI}. Thus, the physical meaning of the barrier
shift remains ambiguous.

\subsection{Spallation reactions}
\label{sec:spal}

Accurate measurements of IMF charge distributions from spallation reactions
below 3~GeV are even more scarce than measurements for fusion
reactions. Moreover, there is some controversy over IMF production cross
sections from 1-GeV p+$^{136}$Xe system; the two existing measurements
\cite{napolitani-xe_xsec,gorbinet-thesis} disagree of about a factor of 3 for
$10\lesssim Z\lesssim30$. This situation prevents us from providing a unique,
predictive parameter set that can simultaneously describe IMF data from fusion
and spallation reactions.

\begin{figure}[t]
  \begin{minipage}[t]{0.47\linewidth}
    \includegraphics[width=\linewidth]{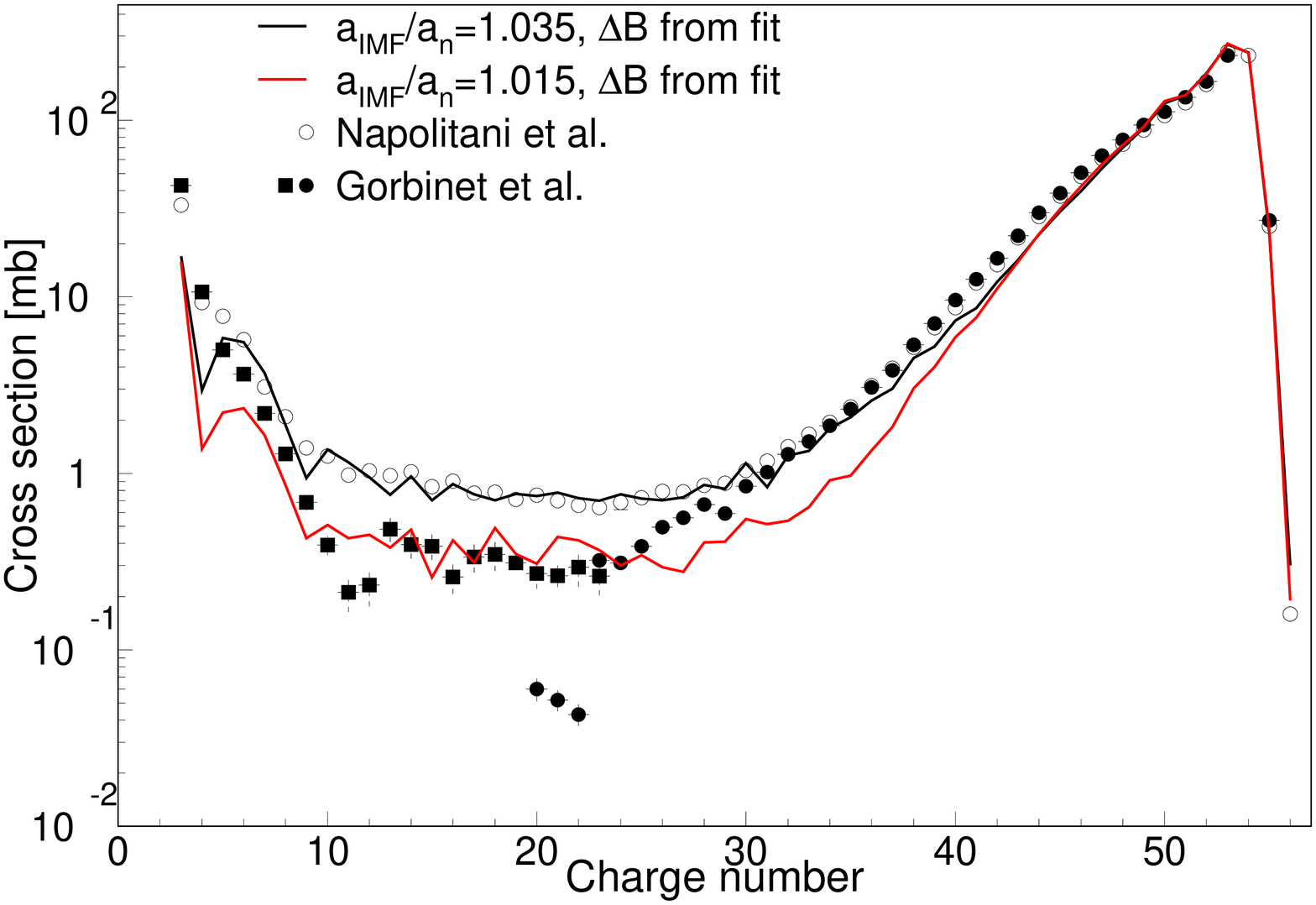}
    \caption{Experimental \protect\cite{napolitani-xe_xsec,gorbinet-thesis} and
      calculated residual charge distributions for 1-GeV p+$^{136}$Xe. The two
      symbols for Gorbinet \etal's data correspond to different techniques (see
      text for more details).}
    \label{fig:xe}
  \end{minipage}\hspace{0.06\linewidth}%
  \begin{minipage}[t]{0.47\linewidth}
    \includegraphics[width=\linewidth]{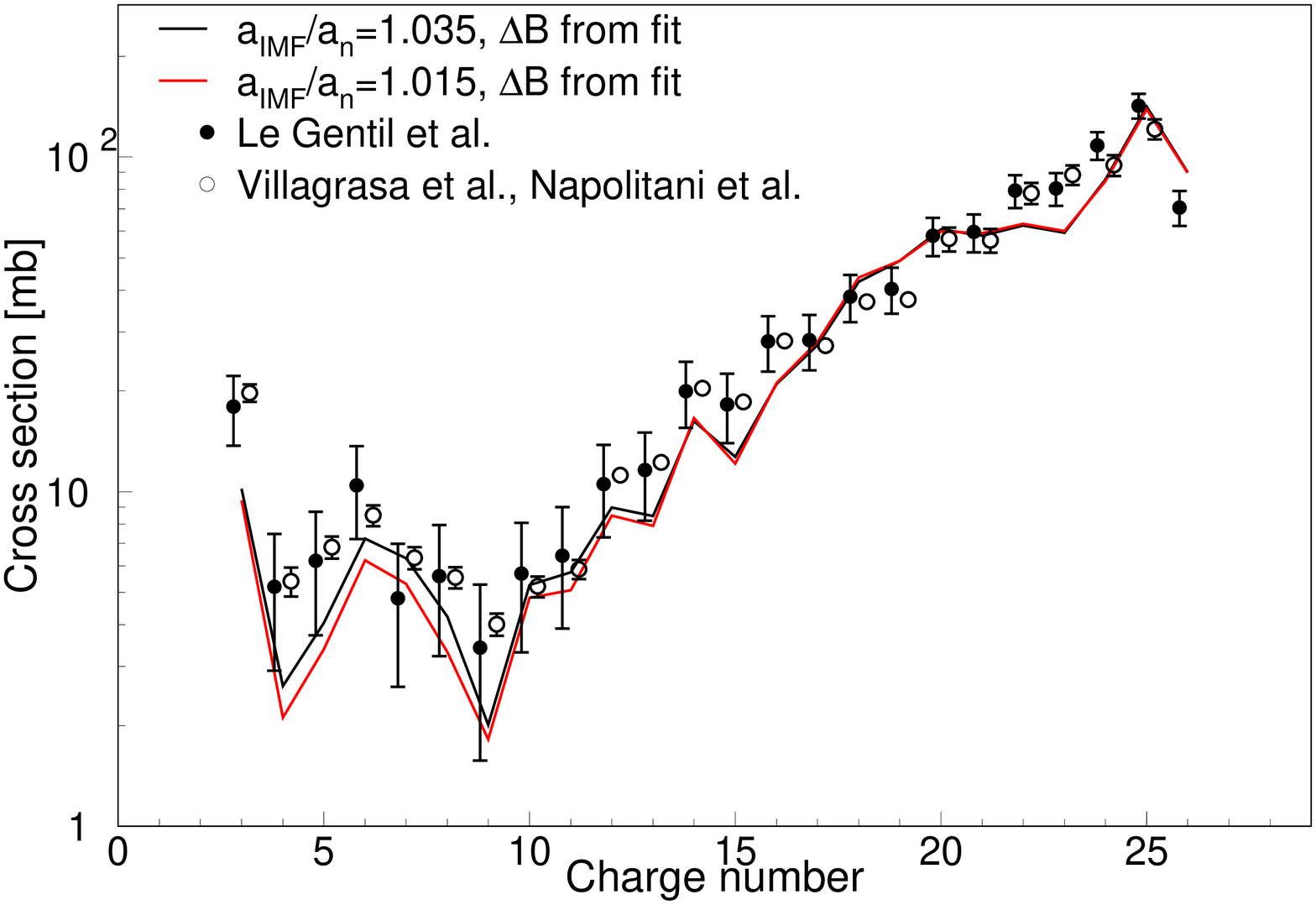}
    \caption{Experimental \protect\cite{villagrasa-fe,napolitani-fe,legentil-fe}
      and calculated residual charge distributions for 1-GeV p+$^{56}$Fe.}
    \label{fig:fe}
  \end{minipage}
\end{figure}

Note that Gorbinet \etal's cross sections in Figure~\ref{fig:xe} actually
consist of two separate data sets. The filled squares for $Z\leq23$ represent
actual fragment-production cross sections. The filled circles for $Z\geq20$
represent cross sections as a function of $Z_\text{max}$, the largest charge
produced in an event. For $Z>27$, the $Z_\text{max}$ cross sections must be
equal to the fragment-production cross sections, since $Z>27$ fragments are
always the largest charges in the events that involve them. For $Z\leq27$, the
actual fragment-production cross sections may be larger than the $Z_\text{max}$
cross sections, if fragments are sometimes produced in coincidence with larger
partners. The two techniques however yield comparable cross sections for $Z=23$,
suggesting that such fragments are never accompanied by larger partners. For
more details, we refer the reader to Gorbinet's Ph.D.\ thesis
\cite{gorbinet-thesis}.

A linear function of $Z^2/A$ was fit to the barrier shifts determined from
fusion reactions (Figure~\ref{fig:shift4}) and applied to all the spallation
reactions. Figure~\ref{fig:xe} shows calculations by \incl/\geminipp for the
1-GeV p+$^{136}$Xe system.

It was mentioned above that IMF charge distributions from fusion reactions are
rather insensitive to the level-density-parameter ratio
$a_\text{IMF}/a_\text{n}$. In spallation reactions, on the contrary, charge
distributions exhibit a larger sensitivity to this parameter, due to the higher
temperatures attained, as illustrated in Figure~\ref{fig:xe}. The model predicts
that fragments with $10<Z<30$ are produced in events with an average excitation
energy $\langle E^*\rangle=411$~MeV, but the distribution extends up to
$\sim750$~MeV. We can then conclude that, as in the case of fission, combining
fusion and spallation data-sets allows us to lift some of the degeneracy of the
model parameters related to IMF production. However, since the available data
sets are in disagreement, we provide two best-fit values for the
$a_\text{IMF}/a_\text{n}$ parameter.

Unfortunately, the 1-GeV p+$^{56}$Fe system does not help to discriminate
between the two candidate values of the $a_\text{IMF}/a_\text{n}$
parameter. Figure~\ref{fig:fe} shows that the IMF cross sections are robust
against variations of $a_\text{IMF}/a_\text{n}$. Note however that the data are
rather accurately described by the model, regardless of the parameter value.

We conclude that IMF production cross sections from fusion and spallation
reactions can be accurately described by introducing a $Z^2/A$-dependent
asymmetric-fission barrier shift and a constant level-density-parameter
ratio. The value of the $a_\text{IMF}/a_\text{n}$ parameter cannot be fixed
until an explanation is found for the 1-GeV p+$^{136}$Xe discrepancy.

\section{Conclusions}

The fusion and spallation entrance channels probe different regions of the
compound-nucleus parameter space and can thus be profitably combined to put
stringent constraints on some of the free parameters of de-excitation models. We
have demonstrated how this strategy can be fruitfully applied to the study of
fission and IMF emission.

In particular, we find that we need to increase Sierk's finite-range barriers
for IMF emission by a few MeV to fit the data. The best-fit barrier shifts
exhibit a phenomenological dependence on $Z^2/A$ of the compound nucleus which
can be fit by a straight line. The description of fission cross sections for
compound nuclei with $A<170$ ($Z^2/A\lesssim30$) would also benefit from a
slight increase in the barriers. It is unclear whether the barrier shift can be
attributed to the deformation dependence of the Wigner energy, because the
best-fit barriers appear to be independent of the $(N-Z)/A$ ratio of the
compound nucleus.

Production of IMF from 1-GeV p+$^{56}$Fe and $^{136}$Xe spallation reactions can
also be accurately described by the same model, provided that one introduces
another free parameter, the ratio of level-density parameters at the conditional
saddle point and in the undeformed configuration
($a_\text{IMF}/a_\text{n}$). Possible values of $a_\text{IMF}/a_\text{n}$ range
from 1.015 to 1.035, depending on the data sets used for the fit.

\providecommand{\newblock}{}


\begin{thebibliography}{10}
\expandafter\ifx\csname url\endcsname\relax
  \def\url#1{{\tt #1}}\fi
\expandafter\ifx\csname urlprefix\endcsname\relax\def\urlprefix{URL }\fi
\providecommand{\eprint}[2][]{\url{#2}}

\bibitem{charity-gemini++}
Charity R~J 2008 {\em Joint ICTP-IAEA Advanced Workshop on Model Codes for
  Spallation Reactions\/} (Trieste, Italy: IAEA) p 139 report INDC(NDC)-0530

\bibitem{sierk-asyBar}
Sierk A~J 1985 {\em Phys.\ Rev.\ Lett.\/} {\bf 55} 582--583

\bibitem{carjan-asyBar}
Carjan N and Alexander J~M 1988 {\em Phys.\ Rev.\ C\/} {\bf 38} 1692--1697

\bibitem{lestone-tilting}
Lestone J~P 1999 {\em Phys.\ Rev.\ C\/} {\bf 59} 1540--1544

\bibitem{rusanov-mass}
Rusanov A~Y, Itkis M~G and Okolovich V~N 1997 {\em Phys.\ Atom.\ Nucl.\/} {\bf
  60} 683--712

\bibitem{moretto-binarydecay}
Moretto L~G 1975 {\em Nucl.\ Phys.\ A\/} {\bf 247} 211--230

\bibitem{hauser-evaporation}
Hauser W and Feshbach H 1952 {\em Phys.\ Rev.\/} {\bf 87} 366--373

\bibitem{gross-friction}
Gross D~H~E and Kalinowski H 1978 {\em Phys.\ Rep.\/} {\bf 45} 175--210

\bibitem{swiatecki-extra_push}
\'{S}wi\k{a}tecki W~J 1980 {\em Prog.\ Part.\ Nucl.\ Phys.\/} {\bf 4} 383--450

\bibitem{boudard-incl}
Boudard A, Cugnon J, Leray S and Volant C 2002 {\em Phys.\ Rev.\ C\/} {\bf 66}
  044615

\bibitem{cugnon-incl45_nd2010}
Cugnon J, Boudard A, Leray S and Mancusi D 2011 {\em J.\ Korean Phys.\ Soc.\/}
  {\bf 59} 955--958

\bibitem{leray-intercomparison}
Leray S, David J~C, Khandaker M, Mank G, Mengoni A, Otsuka N, Filges D,
  Gallmeier F, Konobeyev A and Michel R 2011 {\em J.\ Korean Phys.\ Soc.\/}
  {\bf 59} 791--796

\bibitem{intercomparison-website}
{IAEA} benchmark of spallation models official web site
  \urlprefix\url{http://www-nds.iaea.org/spallations}

\bibitem{pedoux-pions}
Pedoux S and Cugnon J 2011 {\em Nucl.\ Phys.\ A\/} {\bf 866} 16--36

\bibitem{Hinde82}
Hinde D~J, Leigh J~R, Newton J~O, Galster W and Sie S 1982 {\em Nucl.\ Phys.\
  A\/} {\bf 385} 109

\bibitem{Caraley00}
Caraley A~L, Henry B~P, Lestone J~P and Vandenbosch R 2000 {\em Phys.\ Rev.\
  C\/} {\bf 62} 054612

\bibitem{enqvist-lead}
Enqvist T, Wlaz\l{}o W, Armbruster P, Benlliure J, Bernas M, Boudard A,
  Czajkowski S, Legrain R, Leray S, Mustapha B, Pravikoff M, Rejmund F, Schmidt
  K~H, St\'{e}phan C, Ta{\"i}eb J, Tassan-Got L and Volant C 2001 {\em Nucl.\
  Phys.\ A\/} {\bf 686} 481--524

\bibitem{mancusi-gemini++_fission}
Mancusi D, Charity R~J and Cugnon J 2010 {\em Phys.\ Rev.\ C\/} {\bf 82} 044610

\bibitem{fan-se}
Fan T, Jing K, Phair L, Tso K, McMahan M, Hanold K, Wozniak G and Moretto L
  2000 {\em Nucl.\ Phys.\ A\/} {\bf 679} 121--146 ISSN 0375-9474

\bibitem{delis-br75}
Delis D, Blumenfeld Y, Bowman D, Colonna N, Hanold K, Jing K, Justice M, Meng
  J, Peaslee G, Wozniak G and Moretto L 1991 {\em Nucl.\ Phys.\ A\/} {\bf 534}
  403--428 ISSN 0375-9474

\bibitem{jing-mo}
Jing K~X, Moretto L~G, Veeck A~C, Colonna N, Lhenry I, Tso K, Hanold K, Skulski
  W, Sui Q and Wozniak G~J 1999 {\em Nucl.\ Phys.\ A\/} {\bf 645} 203--238

\bibitem{sobotka-decay}
Sobotka L~G, McMahan M~A, McDonald R~J, Signarbieux C, Wozniak G~J, Padgett
  M~L, Gu J~H, Liu Z~H, Yao Z~Q and Moretto L~G 1984 {\em Phys. Rev. Lett.\/}
  {\bf 53}(21) 2004--2007

\bibitem{charity-niobium}
Charity R, McMahan M, Wozniak G, McDonald R, Moretto L, Sarantites D, Sobotka
  L, Guarino G, Pantaleo A, Fiore L, Gobbi A and Hildenbrand K 1988 {\em Nucl.\
  Phys.\ A\/} {\bf 483} 371 -- 405 ISSN 0375-9474

\bibitem{cohen-rldm}
Cohen S, Plasil F and \'{S}wi\k{a}tecki W~J 1974 {\em Ann.\ Phys.\ (N.Y.)\/}
  {\bf 82} 557--596

\bibitem{myers-wigner_energy}
Myers W~D and \'{S}wi\k{a}tecki W~J 1997 {\em Nucl.\ Phys.\ A\/} {\bf 612}
  249--261

\bibitem{napolitani-xe_xsec}
Napolitani P, Schmidt K~H, Tassan-Got L, Armbruster P, Enqvist T, Heinz A,
  Henzl V, Henzlova D, Keli\'{c} A, Pleska\v{c} R, Ricciardi M~V, Schmitt C,
  Yordanov O, Audouin L, Bernas M, Lafriaskh A, Rejmund F, St\'{e}phan C,
  Benlliure J, Casarejos E, {Fernandez Ordonez} M, Pereira J, Boudard A,
  Fernandez B, Leray S, Villagrasa C and Volant C 2007 {\em Phys.\ Rev.\ C\/}
  {\bf 76} 064609

\bibitem{gorbinet-thesis}
Gorbinet T 2011 {\em \'{E}tude des réactions de spallation $^{136}$Xe+p et
  $^{136}$Xe+$^{12}$C à 1 GeV par nucléon auprès de l'accélérateur GSI
  (Darmstadt, Allemagne)\/} Ph.D. thesis Université Paris-Sud 11 Paris, France

\bibitem{villagrasa-fe}
Villagrasa-Canton C {\em et~al.\/} 2007 {\em Phys.\ Rev.\ C\/} {\bf 75} 044603

\bibitem{napolitani-fe}
Napolitani P, Schmidt K~H, Botvina A~S, Rejmund F, Tassan-Got L and Villagrasa
  C 2004 {\em Phys.\ Rev.\ C\/} {\bf 70} 054607

\bibitem{legentil-fe}
{Le Gentil} E {\em et~al.\/} 2008 {\em Phys.\ Rev.\ Lett.\/} {\bf 100} 022701

\end{thebibliography}
\end{document}